# 股票预测：一种基于新闻特征抽取和循环神经网络的方法[*]

张泽亚，陈维政，闫宏飞

北京大学计算机科学与技术系，北京，100871

E-mail: zeyazhang26@gmail.com

**摘　要**：本文提出了一种预测股票涨跌的方法。在特征抽取方面，除了股价信息，我们还提取了与股票相关的新闻特征。我们先依据经验选取了一些能代表新闻利好和利空性质的种子单词，然后基于最优化方法计算出所有单词的利好极性。之后通过单词的利好极性构造出新闻的特征。模型方面，考虑到价格的时序性以及新闻影响的持续性，我们提出了一种循环神经网络模型。在实验中，我们发现相对于基于价格特征的 SVM 分类器，我们提出的方法在股票涨跌预测方面能有超过 5%的提升。

**关键词**：股票预测，特征抽取，循环神经网络

# Stock Prediction: a method based on extraction of news features and recurrent neural networks

Zeya Zhang, Weizheng Chen, Hongfei Yan

Department of Computer Science and Technology, Peking University, Beijing 100871

E-mail: zeyazhang26@gmail.com

**Abstract**: This paper proposed a method for stock prediction. In terms of feature extraction, we extract the features of stock-related news besides stock prices. We first select some seed words based on experience which are the symbols of good news and bad news. Then we propose an optimization method and calculate the positive polar of all words. After that, we construct the features of news based on the positive polar of their words. In consideration of sequential stock prices and continuous news effects, we propose a recurrent neural network model to help predict stock prices. Compared to SVM classifier with price features, we find our proposed method has an over 5% improvement on stock prediction accuracy in experiments.

## 1 引言

股票价格的预测在商业和金融领域具有重要的意义。股票市场的预测在商业界和学术界都受到了广泛的关注。Fama 于 1965 年提出了有效市场假说(Efficient Market Hypothesis)[1]，他认为，股票市场是一个"有效信息"市场，股票价格充分反映了已经发生的事件，以及那些尚未发生但市场预期会发生的事件对股票价格的影响。这一假设为之后的股票预测工作提供了依据。

然而，预测股票价格依然十分困难，因为股票价格受到众多因素的影响。对于单个股票而言，除了国家的货币政策，行业的景气状况等宏观因素，股票上市公司的相关事件等微观因素也会对股票价格产生影响。因此，除了股票自身的价格信息，许多相关工作[2][3][4]中都将股票相关的新闻信息作为预测股票价格的重要依据。

GPC Fung 等在文献[5]中利用实时的新闻信息对股票价格作出预测。他们首先利用线性回归和聚类方法对股票的价格曲线分段，每段时间区间对应价格的上升期和下降期。然后将上升期和下降期内的新闻分别标注为利好消息和利空消息，通过统计方法选择出新闻中的利好和利空特征。最后依据这些新闻中的特征对股票价的涨跌做出预测。该方法忽视了新闻对

于股价影响的持续性。TH Nguyen 等利用主题模型来预测股票价格。在文献[6]中，他们提出了一个融合情感和话题的主题模型，并将该模型运用到股票相关新闻的主题分析中。在获得了每个新闻的主题分布向量后，他们将这个主题分布向量加入到股票预测的特征中，最终获得了不错的预测效果。这种主题模型特征是一种通用的文本特征，忽视了金融市场新闻的特殊性。近几年来，深度学习方法在自然语言处理领域取得了许多进展，Xiao Ding 等将深度学习方法运用到股票预测领域。在文献[7]中，他们提出了一种新的事件抽取方法，从新闻中抽取出结构化的事件。这些结构化的事件成为神经网络的输入，用于预测股票价格。随后，在事件抽取工作的基础上，他们在文献[8]中进一步学习出结构化事件的 event embedding[1]，并使用卷积神经网络模型去预测股票价格。这种模型虽然考虑了事件对于股价的持续影响，但是忽略了多个事件对于股价的综合作用。

除了与股票相关的新闻信息，大众媒体与社交媒体上的内容也被用于股票预测。不过这些媒体上的内容一般不适用于单个股票的预测，只能对股市整体的情况（道琼斯工业指数、上证指数等）作出预测。Johan Bollen 等在文献[9]中运用 Twitter 上的内容对股市的涨跌作出预测。他们使用 OpinionFinder[2] 等工具分析 Twitter 上每天的大众情感，然后将这些情感特征加入到预测模型中，对股市的涨跌作出预测。

本文余下的部分组织如下：第 2 部分给出了问题描述以及本文方法的概览；第 3 部分详细介绍新闻特征抽取和循环神经网络模型；第 4 部分描述了使用的数据集以及模型在数据集上的实验效果；第 5 部分总结本文的内容。

## 2 问题描述和方法综述

股票价格预测是指利用股票价格的历史信息以及与股票相关的市场信息，预测股票在未来一段时间内的涨跌情况或者价格情况。本文的研究主要针对在上海证券交易所[3]挂牌的 A 股[4]股票，也即"上证 A 股"。

### 2.1 股票预测问题

股票在一个交易日会有一些交易特征，例如开盘价，收盘价[5]，最高价，交易量等，其中收盘价一般被用于代表股票在这一天的价格。于是在一段连续的时间内，某只股票会产生一个收盘价格的序列。我们记这段连续的价格序列为：

$$p_1, p_2, \ldots, p_T$$

其中 $p_t$ 代表该股票在第 t 个交易日的收盘价格。

除了基本的价格信息，股票在某些交易日还会有与之相关的新闻事件[6]。我们记这段连续的时间内，与该股票相关的新闻序列为：

$$d_1, d_2, \ldots, d_T$$

其中 $d_t$ 表示第 t 个交易日内发生的新闻集合。新闻文本可以看作单词的序列，对于中文文本而言，即新闻内容进行分词后的序列。新闻集合 $d_t$ 可表示为单词序列：

$$d_t = w_1 w_2 \ldots w_{l_t}$$

$w_i \in V$ 表示单词表 $V$ 中的一个单词，$l_t$ 表示第 t 个交易日新闻的长度。

我们进一步定义股票在每天的涨跌情况为：

---

[1] 类似于 word2vec 中 word embbeding 是对单词的表示，event embedding 是对结构化事件的表示
[2] OpinionFinder 网站：http://mpqa.cs.pitt.edu/opinionfinder/，一个开源情感极性分析工具
[3] 上海证券交易所官网：www.sse.com.cn
[4] A 股：人民币普通股票，它是由我国境内的公司发行，供境内机构、组织和个人以人民币认购和交易的普通股票。
[5] 指股票在交易日里最后一笔买卖的成交价格
[6] 与股票相关的新闻事件：上市公司公告、收购、内幕丑闻等事件

$$c_1, c_2, \ldots, c_T$$

其中$c_t$表示第 t 个交易日内股票价格的涨跌情况，1 表示价格上涨，0 表示价格下跌或者不变：

$$c_t = \begin{cases} 1, & if\ p_t > p_{t-1} \\ 0, & if\ p_t \leq p_{t-1} \end{cases} \tag{1}$$

股票价格的涨跌预测问题可定义如下：

已知一段连续交易日内某只股票的收盘价格序列和与之相关的新闻序列：

$$p_1, p_2, \ldots, p_t$$
$$d_1, d_2, \ldots, d_t$$

预测该股票在下一个交易日价格是上涨还是下跌，即预测$c_{t+1}$。 (2)

## 2.2 预测方法总览

本文提出了一种基于单词点互信息的新闻特征抽取方法,然后将该特征运用到股票价格的预测中，并提出了一种基于循环神经网络[7]的股票预测模型。预测方法的流程如图 1 所示。

特征抽取方面，首先选取利好关键词和利空关键词的种子集合，然后使用最优化方法选出利好与利空的标准关键词集合，并用标准关键词集合计算其他单词的利好极性。之后，基于单词的利好极性构造出新闻文本的利好利空特征。

股票预测模型方面，使用循环神经网络有着直观的考虑。首先股票价格在短期内有着一定的时序相关性，其次新闻事件对于股票价格的影响具有持续性，即新闻事件会在一段时间内对股票价格产生持续的影响。

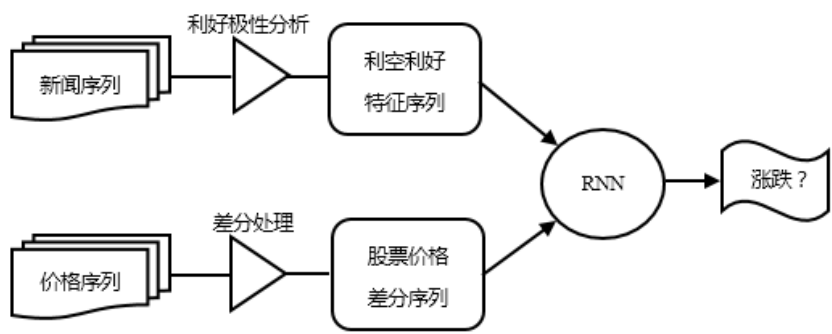

图 1：股票价格涨跌预测流程图

Fig 2：flow chart of stock price prediction

# 3 特征和模型

股票涨跌预测问题可以看作一个二分类问题。股票预测问题的输入中包含两种不同类型的数据，即新闻文本数据和股票价格数值数据。由于新闻文本数据不能直接作为分类器的输入，因此人们一般对新闻文本进行一些处理，抽取出文本中的特征作为输入。

## 3.1 单词的利好极性分析

新闻对于股票的影响可分为利好和利空两类。利好新闻是指推动股价上升的消息，比如中央银行降息降准，上市公司收购其他公司等。利空新闻是指促使股价下降的消息，比如上市公司经营业绩恶化，银行利率调高等。

为了分析一则新闻是利好新闻还是利空新闻，本文首先分析新闻中单词的利好极性，然后基于单词的利好极性构造出新闻的利好利空特征，将这一特征作为分类模型的输入。

---

[7] 循环神经网络，即 Recurrent Neural Network(RNN)

假设我们有一个足够大的与股票相关的新闻文档集合 D：

$$D = \{doc_1, doc_2, \ldots, doc_N\} \quad (3)$$

每个新闻可以看作一个单词的序列：

$$doc_i = w_1 w_2 \ldots w_{n_i} \quad (4)$$

单词可以看作在文档集上的一个分布，有些单词会更多地在利好新闻中出现，有些单词会更多地在利空新闻中出现，还有些单词则均匀地分布在利好和利空新闻中。单词 w 在文档集中出现的概率记为 p(w)，两个单词 w 和 v 之间的点互信息[10]记为 pmi(w,v)：

$$p(w) = \frac{|\{i | w \in doc_i\}|}{N} \quad (5)$$

$$pmi(w, v) = \ln \frac{p(w,v)}{p(w)p(v)} = \ln \frac{N * |\{i | w \in doc_i \text{ and } v \in doc_i\}|}{|\{i | w \in doc_i\}| * |\{i | v \in doc_i\}|} \quad (6)$$

点互信息描述了两个单词的相关性。文本中相互关联的单词之间的点互信息会比较大，比如单词"利好"和"升值"，而在文本中没有什么联系的单词之间的点互信息会比较小，比如单词 "利好"和"养殖"。

假设我们通过金融市场经验获取了一些利好关键词和利空关键词，其中利好关键词是指更有可能在利好新闻中出现的单词，利空关键词是指更有可能在利空新闻中出现的单词。我们把他们分别称为利好种子集 $P_{seed}$ 和利空种子集 $N_{seed}$：

$$P_{seed} = \{u_1, u_2, \ldots, u_{|P_{seed}|}\} \quad (7)$$

$$N_{seed} = \{v_1, v_2, \ldots, v_{|N_{seed}|}\} \quad (8)$$

我们希望找到一组新的标准单词将利好种子集和利空种子集尽可能地区别开来。为此，我们定义标准利好集 P 和标准利空集 N，以及单词 w 在 P 和 N 上的利好极性 polar(w)：

$$P_{std} = \{w_{p1}, w_{p2}, \ldots, w_{pK}\} \quad (9)$$

$$N_{std} = \{w_{n1}, w_{n2}, \ldots, w_{nK}\} \quad (10)$$

$$polar(w) \triangleq \frac{1}{K} \sum_{v \in P_{std}} pmi(w,v) - \frac{1}{K} \sum_{v \in N_{std}} pmi(w,v) \quad (11)$$

其中 $P_{std}$ 为一组数量为 K 的未知的利好标准单词，$N_{std}$ 为一组数量为 K 的未知的利空标准单词。它们分别与利好新闻与利空新闻相关联，用于计算其他单词的利好极性。单词 w 的利好极性 polar(w)表示了单词在利好新闻中出现的概率；利好极性越大，单词在利好新闻中出现的概率越大，在利空新闻中出现的概率越小。

为了找到一组好的利好标准集与利空标准集，我们定义如下的最优化问题：

已知文档集合 D，利好种子集 $P_{seed}$ 和利空种子集 $N_{seed}$，求解最优标准集 $P^*$ 和 $N^*$：

$$P^*, N^* = \underset{P_{std}, N_{std}}{\mathrm{argmax}} \left[ \frac{1}{|P_{seed}|} \sum_{w \in P_{seed}} polar(w) - \frac{1}{|N_{seed}|} \sum_{w \in N_{seed}} polar(w) \right],$$

$$\text{s.t. } |P_{std}| = |N_{std}| = K \quad (12)$$

为了求解最优标准集 $P^*$ 和 $N^*$，我们定义单词在种子集合上的极性为：

$$polar_{seed}(w) \triangleq \frac{1}{|P_{seed}|} \sum_{v \in P_{seed}} pmi(w,v) - \frac{1}{|N_{seed}|} \sum_{v \in N_{seed}} pmi(w,v) \quad (13)$$

综合式子(11),(12)和(13)，不难发现：

$$P^*, N^* = \underset{P_{std}, N_{std}}{\operatorname{argmax}} \left[ \frac{1}{K} \sum_{w \in P_{std}} polar_{seed}(w) - \frac{1}{K} \sum_{w \in N_{std}} polar_{seed}(w) \right],$$
$$\text{s.t. } |P_{std}| = |N_{std}| = K \tag{14}$$

由于种子集合$P_{seed}$和$N_{seed}$是已知的，因此对于单词表 V 中的任意单词 w, $polar_{seed}(w)$是确定的。如果我们把所有的 M 个单词按照单词的$polar_{seed}(w)$从大到小排序，即：

$$polar_{seed}(w_{s1}) > polar_{seed}(w_{s2}) > polar_{seed}(w_{s3}) > \cdots > polar_{seed}(w_{sM}) \tag{15}$$

那么最优标准利好集$P^*$为序列(15)中的前 K 个单词，最优标准利空集$N^*$为序列(15)中的后 K 个单词，即：

$$P^* = \{w_{s1}, w_{s2}, \ldots, w_{sK}\} \tag{16}$$
$$N^* = \{w_{s(M-K+1)}, w_{s(M-K+2)}, \ldots, w_{sM}\} \tag{17}$$

在获得最优标准利好集$P^*$和最优标准利空集$N^*$后，我们最终可以定义单词在$P^*$和$N^*$上的利好极性，作为后续新闻特征抽取的基础：

$$\operatorname{polar}(w) \triangleq \frac{1}{K} \sum_{v \in P^*} pmi(w,v) - \frac{1}{K} \sum_{v \in N^*} pmi(w,v) \tag{18}$$

### 3.2 新闻的特征构造

抽取新闻特征有多种方法。One-Hot 是一种简单的文本表示方法，它将新闻看作|V|维向量，向量的每一维对应词汇表 V 上的一个单词。One-Hot 表示方法的缺点在于特征向量比较稀疏，维度非常大。情感分析特征也可以作为新闻的特征[11][12]，但是这种在文本上的通用情感特征忽视了股票预测领域的特殊性。

在 3.1 节中，我们在最优标准集$P^*$和$N^*$上定义了所有单词的利好极性(18)，在此基础上，可以构造新闻特征。综合考虑新闻的维度大小以及新闻特征的表示能力，我们将利好极性数值范围划分为若干段，然后将每段数值对应的单词捆绑,以此表示每一篇新闻。具体描述如下：

假设单词表 V 中单词的最大利好极性为 Max，最小利好极性为 Min，我们希望将利好极性数值范围分为 L 段，每段对应一个利好极性区间 I：

$$\operatorname{Max} = \max_{w \in V} polar(w) \tag{19}$$

$$\operatorname{Min} = \min_{w \in V} polar(w) \tag{20}$$

$$I_j = \left[ Min + (j-1) * \frac{Max-Min}{L}, Min + j * \frac{Max-Min}{L} \right], j \in \{1, 2, \ldots, L\} \tag{21}$$

对于第 i 个交易日的新闻$doc_i = w_1 w_2 \ldots w_{n_i}$，它的特征$f(doc_i)$为：

$$f(doc_i) = (x_{i1}, x_{i2}, \ldots, x_{iL})^T \tag{22}$$

其中$x_{ij} = \frac{1}{n_i} |\{k | polar(w_k) \in I_j\}|$

式子(22)中定义的新闻特征是新闻中的单词在利好极性维度上的分布的直方图近似，直方图的区间数量为 L。这种特征可以通过调节 L 的大小来控制特征维度的大小，并且直观地反映了新闻对于股票价格的利好利空影响。我们将它作为后续预测模型的输入，如图 1 所示。

### 3.3 股票价格的差分处理

除了与股票相关的新闻信息，股票价格自身的历史信息也很重要。相对于股价的绝对数

值，股价的涨跌幅对于预测股价的涨跌更加有效。对于不同的股票，股价的绝对数值差异很大，但是股价在单个交易日内的涨跌幅都在-10%到 10%之间。在实验中，我们使用的价格特征是股价在每个交易相对前一个交易日的涨跌幅。

假设股票价格在一段连续的交易日内的价格序列为：

$$p_1, p_2, \ldots, p_T$$

那么股价在第 i 个交易日的涨跌幅$r_i$为：

$$r_i = \frac{p_i - p_{i-1}}{p_{i-1}}$$

由此我们获得了股价在这段交易日内的差分序列（涨跌幅序列）：

$$r_1, r_2, \ldots, r_T \tag{23}$$

式子(23)中定义的股价差分序列被用于后续预测模型的输入，如图 1 所示。

### 3.4 循环神经网络预测模型

从股票价格的涨跌预测问题(2)中可以发现，不论是股价信息还是新闻信息，都是和时间相关的序列。相对于其他的分类器（支持向量机[13]，决策树[14]等），循环神经网络能够有效地描述特征在时序上的影响[15]。在股票预测问题中，一则新闻对于股价的影响往往不局限于一个交易日，而是在一段交易日内。本文中的循环神经网络预测模型如图 2 所示。

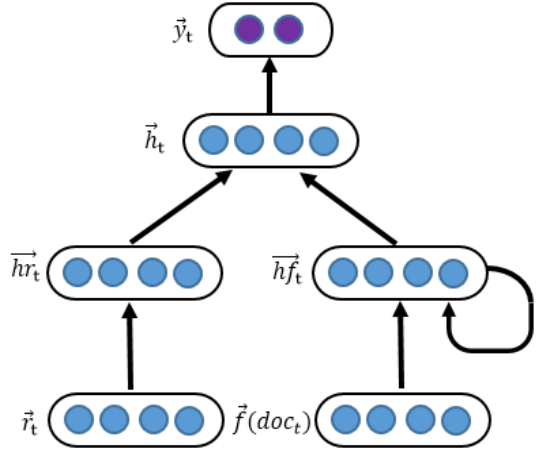

图 3：循环神经网络预测模型

Fig 4: Recurrent Neural Network model

神经网络的输入层包含两个部分，一部分是式子(23)中定义的股价差分序列$r_1, r_2, \ldots, r_T$，该序列利用上一个交易日股价涨跌信息预测下一个交易日的涨跌，体现了价格信息对股票预测的作用。另一部分是和新闻相关的特征序列，即式子(22)中定义的$f(doc_t)$, $t \in \{1, 2, \ldots, T\}$。这部分输入表示了新闻特征对股价的影响。

神经网络的第一个隐含层包含两部分，分别对应于输入层的两个部分。第一部分$\overrightarrow{hr_t}$由$\overrightarrow{r_t}$经过变换得到，表示第 t 个交易日价格信息的隐含向量。第二部分$\overrightarrow{hf_t}$表示前 t 个交易日新闻信息的综合隐含向量，它由当前交易日的新闻特征$\vec{f}(doc_t)$，以及上一个交易日的综合隐含向量$\overrightarrow{hf_{t-1}}$经过叠加变换得到。通过将上一个交易日的新闻特征保留到下一个交易日，预测模型能够有效地体现新闻信息对于股价的持续性影响。

神经网络的第二个隐含层对第一个隐含层的信息进行综合，将价格信息$\overrightarrow{hf_t}$和新闻信息$\overrightarrow{hf_{t-1}}$融合，形成一个综合考虑了价格与新闻的特征向量。

神经网络的输出层$\vec{y_t}$包含两个单元，分别对应股价在下个交易日上涨和下跌的概率。

模型的隐含层使用的非线性变换为ReLU函数[8]，输出层使用的变换是Softmax函数[9]。循环神经网络的预测过程如表1所示。

表 1:循环神经网络的预测
Tab 1: prediction process of recurrent neural network

| 循环神经网络的预测 |
| --- |
| 输入：价格差分序列$r_1, r_2, ..., r_t$，新闻特征序列$f(doc_1), f(doc_2), ..., f(doc_t)$，模型参数矩阵$W_r, W_f, W_{hr}, W_{hf}, V_{hf}, W_h$，参数向量$b_{hr}, b_{hf}, b_h$<br>输出：第 t+1 个交易日的涨跌类别和上涨概率$<\hat{c}_{t+1}, \hat{y}_{t+1}>$<br>1.　　$hf_0 = \vec{0}$<br>2.　　**for** i from 1 to t:<br>3.　　　　$hr_i := ReLU(W_r \cdot r_i + b_{hr})$<br>4.　　　　$hf_i := ReLU(W_f \cdot f(doc_i) + V_{hf} \cdot hf_{i-1} + b_{hf})$<br>5.　　　　$h_i := ReLU(W_{hr} \cdot hr_i + W_{hf} \cdot hf_i + b_h)$<br>6.　　　　$y_i := Softmax(W_h \cdot h_i)$<br>7.　　**endfor**<br>8.　　**if**　$y_t[1] > y_t[0]$:<br>9.　　　　return $<1, y_t[1]>$<br>10.　**else**<br>11.　　　return $<0, y_t[1]>$<br>12.　**endif** |

## 3.5 模型的参数学习

为了获得模型中的参数，我们需要定义模型的损失函数。

假设在一段长度为 T+1 的交易日内，观测到某只股票的价格序列为$\vec{p} = p_1, p_2, ..., p_{T+1}$，股票对应的新闻序列为$\vec{d} = d_1, d_2, ..., d_T$，图 2 中的预测模型的参数为$\vec{\theta}$。实验中使用的损失函数如式子(24)所示。

$$L(\vec{p}, \vec{d}; \vec{\theta}) = -\sum_{i=1}^{T}[c_{i+1} \ln \hat{y}_{i+1} + (1-c_{i+1})\ln(1-\hat{y}_{i+1})] + \lambda \|\theta\|_F^2 \qquad (24)$$

根据式子(1)和式子(23)，由价格序列$\vec{p}$可以获得价格的涨跌观测序列$\vec{c} = c_1, c_2, ..., c_{T+1}$，以及股票的价格差分序列$\vec{r} = r_1, r_2, ..., r_T$。根据式子(22)，由股票的新闻序列$\vec{d}$可以计算出新闻的特征序列$\vec{f} = f(doc_1), f(doc_2), ..., f(doc_T)$。价格的差分序列$\vec{r}$和新闻的特征序列$\vec{f}$是图 1 中模型的输入，$\vec{\theta} = \{W_r, W_f, W_{hr}, W_{hf}, V_{hf}, W_h, b_{hr}, b_{hf}, b_h\}$是模型的参数。如表 1 所示，预测模型会预测股价从第 2 到第 T+1 个交易日中每天的涨跌类别$\hat{c_t}$以及价格上涨的概率$\hat{y_t}$。

式子(24)中的损失函数包含两个部分，前一部分是交叉熵损失函数，表示模型预测错误的代价，后一部分是 L2 正则化项，代表对模型自身复杂度的惩罚。参数$\lambda$是模型的超参数，表示模型正则化部分的比重。

参数的学习采用随机梯度下降(SGD)[16]的方法。实验中循环神经网络的代码实现以及梯度的具体计算使用了 Python 中的 Theano[10]库。

---

[8] ReLU，即 Rectified Linear Units，ReLU(x)=max(0,x)
[9] Softmax 函数将 K 维向量 x 变换为 K 维向量 Softmax(x),其中$Softmax(x)_j = e^{x_j}/\sum_i e^{x_i}$,j = 1,2, ..., K
[10] Theano 库是 Python 中一个支持数学表达式的符号计算的库,网址为 http://deeplearning.net/software/theano/

# 4 实验

## 4.1 数据集描述

数据集分为股票价格和新闻语料两部分。我们从投资之家[11]上抓取了上证 A 股市场从 2013 年 1 月到 2015 年 10 月的股票价格信息,从搜狐证券版块[12]获取了这段时间内的新闻页面。为了将新闻与股票对齐,我们保留标题中出现股票名称的新闻,并认为这则新闻与该只股票相关。对于标题中没有出现股票名称的新闻,我们将其过滤。最后新闻集中包含了 85060 篇与上证 A 股市场相关的新闻。数据集的基本信息如表 2 所示。

表 2:数据集概况
Tab 2: Overview of dataset

| 交易日起止时间 | 2013-01-01 至 2015-08-31 |
|---|---|
| 交易天数 | 644 |
| 股票数量 | 1078 |
| 相关新闻篇数 | 85060 |

数据集包含了上证 A 股市场 1000 多只股票从 2013 年 1 月到 2015 年 8 月的价格信息。每个交易日都有开盘价、收盘价、最高价和交易量等信息。实验中我们按照股票相关新闻的数量,选取了两组股票,分别对它们进行预测。如表 3 所示,前 5 只股票是相关新闻数量较多的股票,后 5 只股票是相关新闻数量较少的股票。

表 3:新闻密度较高的股票
Tab 3: stocks with lots of related news

| 股票代码 | 股票名称 | 交易日数 | 相关新闻天数 |
|---|---|---|---|
| 600030 | 中信证券 | 644 | 304 |
| 600519 | 贵州茅台 | 644 | 198 |
| 601998 | 中信银行 | 644 | 261 |
| 601318 | 中国平安 | 643 | 229 |
| 601628 | 中国人寿 | 643 | 205 |
| 600795 | 国电电力 | 623 | 109 |
| 601186 | 中国铁建 | 613 | 94 |
| 600277 | 亿利能源 | 556 | 73 |
| 600887 | 伊利股份 | 619 | 156 |
| 600703 | 三安光电 | 584 | 112 |

表 3 中前 5 只股票在搜狐证券中有较多的相关新闻,比如"中信证券", 644 个交易日中有 304 天出现了与之相关的新闻,平均每两个交易日有一则相关新闻。而后 5 只股票相关新闻数量较少,比如"中国铁建",613 个交易日中只有 94 天出现相关新闻。

表 4:利好种子集与利空种子集
Tab 4: Good and bad set of seed words

| 种子集 | 内容 |
|---|---|
| 利好种子集 | {受益、提升、改善、稳健、看好、有望、增、收购、利好、优势} |
| 利空种子集 | {下滑、低于、下降、拖累、跌、降、亏损、违规、处罚、利空} |

---

[11] http://www.inv.org.cn/
[12] http://stock.sohu.com/

特征抽取方面，为了获得每个单词的利好极性，需要通过经验选取式子(7)(8)中的利好种子集以及利空种子集。实验中选取的种子集如表 4 所示。容易发现，利好种子集中单词是常用的对上市公司进行正面评价和预期的词汇，而利空种子集中的内容包含了对上市公司的负面评论。

**4.2 实验结果**

为了检验论文中提出的新闻特征的效果，以及 RNN 预测模型的作用，我们设计了几组不同的特征和模型。第一种方法使用股票价格作为特征，使用 SVM 分类器作为预测模型（记为 Price+SVM）。这种方法使用前 3 个交易日的涨跌幅度作为特征，使用 SVM 分类器预测下个交易日股价的涨跌情况。第二种方法使用股票价格以及新闻序列作为特征，使用 SVM 分类器作为预测模型（记为 Price+News+SVM）。该方法使用前 3 个交易日的涨跌幅度，以及当天的新闻作为特征，以 SVM 分类器作为模型，预测下一个交易日股价的涨跌。第三种方法使用股票价格以及新闻序列作为特征，使用 RNN 模型作为预测方法（记为 Price+News+RNN）。

对于每只股票，数据集有 644 个交易日的价格信息和新闻信息。我们把较早的 80%的交易日作为训练数据，剩余的 20%的交易日作为测试数据。我们用预测模型的分类准确率作为判断模型优劣的标准。我们按照股票相关新闻数量，将股票分为两组，便于比较模型在不同新闻密度的股票上的效果。

三种不同的方法在新闻密度较高的 5 只股票上的结果如表 5 所示，在新闻密度较低的 5 中股票上的结果如表 6 所示。

表 5：高新闻密度股票上不同模型的准确率
Tab 5: Accuracy of different models on stocks with lots of news

| 股票名称 | Price+SVM | Price+News+SVM | Price+News+RNN |
| --- | --- | --- | --- |
| 中信证券 | 0.512 | 0.56 | 0.6 |
| 贵州茅台 | 0.488 | 0.56 | 0.592 |
| 中信银行 | 0.496 | 0.504 | 0.52 |
| 中国平安 | 0.504 | 0.528 | 0.544 |
| 中国人寿 | 0.52 | 0.528 | 0.536 |
| 平均 | **0.504** | **0.536** | **0.5584** |

表 6：低新闻密度股票上不同模型的准确率
Tab 6: Accuracy of different models on stocks with few news

| 股票名称 | Price+SVM | Price+News+SVM | Price+News+RNN |
| --- | --- | --- | --- |
| 国电电力 | 0.488 | 0.496 | 0.496 |
| 中国铁建 | 0.52 | 0.512 | 0.528 |
| 亿利能源 | 0.491 | 0.491 | 0.5 |
| 伊利股份 | 0.484 | 0.532 | 0.548 |
| 三安光电 | 0.496 | 0.513 | 0.538 |
| 平均 | **0.4958** | **0.5088** | **0.522** |

从表 5 中可以发现，方法一(Price+SVM)使用价格特征作为唯一的特征来预测股价涨跌，在 5 只股票上的平均准确率为 50.4%，与随机预测相比几乎没有提升。第二种方法(Price+News+SVM)除了价格信息，还将新闻信息融入到特征中，并使用 SVM 分类器进行预测。其平均准确率为 53.6%，相对于方法一提升了 3.2 个百分点，这说明了文中提出的新闻特征对于股票预测有一定的作用。方法三(Price+News+RNN)与方法二在特征选择方面相同，

在模型选择方面，方法三使用文中提出的 RNN 模型，而方法二使用 SVM 分类器。从效果来看，方法三的平均准确率为 55.84%，比方法二高出 2.24 个百分点。

在表 6 中，方法一(Price+SVM)在 5 只股票上的平均准确率为 49.6%。而方法二和方法三的平均准确率分别为 50.9%和 52.2%，相对于方法一的结果，分别有 1.3%和 2.6%的提升。综合分析表 5 和表 6 的结果，新闻密度对于股票预测的效果有影响。股票的相关新闻密度越高，文中提出的新闻特征以及预测模型的预测效果越明显。

## 5  总结

本文在股票预测方面，提出了一种基于点互信息分析的新闻特征，以及一种将价格特征和新闻特征相结合的循环神经网络预测模型。实验中，我们发现在相关新闻数量较多的股票上，文中提出的新闻特征能帮助提升 3.2%的准确率，而文中提出的预测方法能提升 5.44%的准确率。